Local probability model for the Bell correlation based on the statistics of chaotic light and non-commutative processes

Louis Sica

lousica@aol.com

Chapman University, Orange, CA 92866

and

Inspire Institute, Inc., Alexandria, VA 22303

Abstract

As discussed below, Bell's inequalities and experimental results rule out commutative hidden variable models as a basis for Bell correlations, but not necessarily non-commutative probability models. A local probability model is constructed for Bell correlations based on non-commutative operations involving polarizers. As in the entanglement model, the Bell correlation is obtained from a probability calculus without explicit use of deterministic hidden variables. The probability calculus used is associated with chaotic light. Joint wave intensity correlations at spatially separated polarization analyzers are computed using common information originating at the source. When interpreted as photon count rates, these yield quantum mechanical joint probabilities after the contribution of indeterminate numbers of photon pairs greater than one is subtracted out. The formalism appears to give a local account of Bell correlations.

# 1   Introduction

The goal of this paper is to show that a local probability model may account for the Bell cosine correlation in place of the non-local quantum mechanical probability model based on entanglement and the interaction of spatially separated beams. The local probability model constructed is based on the quantum optical statistics of spontaneous-parametric-down-conversion (SPDC), and mixed Poisson and binomial processes. These statistical processes, ordinarily invoked to describe the properties of chaotic light, determine the correlation of wave intensities that are in turn interpreted in terms of photon counts. The resulting Bell correlation of intensities, or count rates, is obtained from accepted statistical facts alone, without further definition of individual photons or their relation to waves.

In the paragraphs below, the author explains his motivation for undertaking this work based on the deduction from Bell's inequalities and experimental results that local commutative hidden variables are ruled out as a basis for Bell correlations, but not hidden variables arising from local non-commutative operations. The operations of polarization analyzers on electromagnetic fields as used in Bell experiments are intrinsically non-commutative, so the approach taken evades a crucial logical barrier to the construction of alternatives to the accepted model.

A majority of the quantum optics community accepts the algebra of entanglement and associated probability calculus as the explanation for Bell correlations. Nonetheless, interest remains high in the question of whether local hidden variables may ultimately account for them [1]. The success of the usual quantum entanglement probability calculation in accounting for Bell correlations (see Shih [2] for a review) is marred by its seemingly unavoidable interpretation in terms of instantaneous action at a distance. In the quantum theory of an entangled state, a measurement carried out on a particle at one location instantaneously changes the quantum state of a second distant particle independently of the magnitude of their spatial separation. Thus, if the quantum mechanical description of the situation is complete, the fundamental change in quantum state of the second particle implies a change in its physical state, and this implies instantaneous causation reaching across the universe. Thus reasoned Einstein in his letters, according to Fine [3]. While some have argued that a conflict with relativity does not exist since humans cannot send information faster than light using Bell correlations, not everyone has found this argument persuasive. Shimony has indicated his agreement with Bell that this explanation does not satisfactorily resolve the issue [4]. The author would judge the issue as being not whether man can signal instantaneously at a distance using quantum correlations, but whether nature can act instantaneously at a distance to change a physical state.

There is an additional closely related issue: using the entangled state to compute the Bell correlation away from its maximum at equal detector settings, the interference of spatially separated light beams is implied [5]. This is a more dramatic assumption of non-locality than that occurring at equal detector settings.

The question raised in this paper is this: is the entanglement explanation necessary to account for quantum correlations, or merely sufficient? Based on the model described below, it appears that entanglement is sufficient, but not necessary. It is the common wisdom that a local hidden variable explanation of the Bell correlation is ruled out by experimental violation of Bell's inequalities and the associated logic that constitutes Bell's theorem. However, flaws in the logic of the Bell theorem disallow this conclusion. The historical reasoning used in deriving the Bell inequalities and theorem is based on the unstated assumption of a commutative hidden variable model that, as will be shown immediately below, does not hold even for basic non-commutative classical measurement operations performed on polarized light.

Given a linearly polarized beam of light, an ideal linear polarizer with transmission direction oriented at 45° to the initial beam polarization direction would pass 50% of the light intensity. A second polarizer with transmission direction at 90° to the initial bean polarization direction would transmit 50% of the remaining intensity. But if the polarizer settings were interchanged, the intensity of light transmitted by each would be zero. (In a quantum mechanical description, the percentages would indicate probabilities.) This shows that in general, one may not construct a table of measurement output predictions, even for variables that behave classically, independently of a specification of the sequence of measurement operations. If the measurement outcomes depend on random processes, predictions in general will involve conditional probabilities. (Counterfactual measurements, if considered, are also probabilistically conditional on the outcomes of actual measurements.) Unfortunately, while Bell raised profound questions in his considerations of hidden variables, the model he used in deriving the Bell inequalities implicitly assumed that all hidden variables are commutative, and thus ignored the fact that many classical as well as quantum mechanical measurement procedures are non-commutative [6]. Recently, Christian [7], reached similar conclusions regarding this critical flaw in the derivation of Bell's theorem.

Bell's assumption of commutativity in the derivation of his inequality is contained in his use of [8] three measurement outputs from a response function defined at all output angles. This, and the assumption that ensemble average correlations depend on the differences of angular settings, effectively define a stochastic process that is spatially stationary in the wide sense [9] in angle variables. These assumptions, if consistent with the characteristics of polarization operations on polarized light, would justify the fact that in no Bell optical experiment using polarization analyzers known to the author, have data sets ever been cross correlated as the derivation of the Bell inequalities prescribes. If the underlying process were truly wide sense stationary this would not matter, since the Bell inequality would then be satisfied with all correlations having the same functional form. The fact that the inequality is not satisfied with all correlations having the Bell co-sinusoidal form indicates that the process producing the correlations is not wide sense stationary and therefore not commutative.

Bell also made an explicit assumption of locality, along with the unstated assumptions just indicated. However, it has not been realized until relatively recently that the inequality that he derived holds independently of the restrictive assumptions on which it has been believed to rest. The assumption of locality, and the unstated assumption of commutativity embodied in the assumption of a wide sense stationary process model are sufficient, but not necessary to derive the Bell inequality. It is easy to show that such inequalities (in either three or four variables) are satisfied by the cross correlations of any data sets consisting exclusively of ± 1's independently of statistical or probability considerations, by merely assuming that the appropriate number of data sets exist [10]. This fact implies that a rethinking of the consequences of Bell inequality violation is in order.

It is apparent, since the recent significant contributions of Hess [11] and De Raedt [12], that the basic mathematical facts regarding Bell inequalities in the probability domain have been independently rediscovered several times over the years beginning with Boole [13] in 1862. In the data correlation domain originally considered by Bell, Eberhard [14] derived the Bell inequalities from imagined data without hidden variables. For the present author [10], this is the preferred approach, and it leads to immediate logical consequences: the Bell inequality holds with mathematical generality, and independently of physical assumptions. For the non-commutative variables that occur in Bell experiments, simultaneous cross-correlations among the data sets cannot all have a cosine of angular difference form, since that violates the inequality that all data identically satisfy. Thus, any stochastic process underlying the Bell correlations cannot be defined entirely in terms of commutative operations, since that implies that all correlations are stationary and have the same mathematical form.

The above reasoning is confirmed by the observation that when consistent simulated quantum mechanical probabilities are used to compute correlations among the assorted variables that Bell used, experimentally measurable, and non-commuting counterfactuals, it is found that correlations involving counter-factuals have a form different from the familiar Bell cosine of angular differences [8]. The resulting set of correlations satisfies the Bell inequalities. This is true even though non-local information is used in the models.

While by definition, correlations involving counterfactuals are not measurable, modifications of the data processing used in multiple Bell experiments allow cross-correlation of measurable data as required by the derivation of Bell inequalities. The additional correlations are conditional on the primary ones, and their predicted forms are different from the usual Bell correlations [15]. Again, the Bell inequalities are satisfied.

Recently, Adenier and Khrennikov [16] analyzed data from classic Bell correlation experiments and found anomalies in the magnitudes of some joint probabilities, although the observed effects cancel out to produce the Bell cosine correlation. They found that Aspect's thesis reported the observation of similar anomalies in some data of his classic experiment. Interpretation of these anomalies is not yet settled, but ranges from rejection of the fair sampling assumption to rejection of the condition of entanglement. Thus, experimental data can and do exist that yield the proper Bell correlation, yet have other characteristics inconsistent with the usual entanglement explanation.

In addition to the purely logical difficulties in the derivation and use of Bell inequalities, there is the empirical difficulty that all experiments performed to date require the use of the fair sampling assumption to obtain agreement with the computed theoretical visibility of the Bell correlation. This means that the fact that 90-95% of the data consists of singles counts (counts occurring on one side or the other of the apparatus but not both) is assumed to be unrelated to the cause of the correlations, and so may be ignored in the computation of visibility. But can we be sure that the high singles counts are really statistically innocuous when local hidden variables have been shown to provide a basis for the outcomes of the GHZ theorem if strategically located zeros are placed among the data [17]?

The issues raised above provide justification for seeking alternative explanations for Bell correlations, in spite of the widespread belief that a novel kind of nonlocal effect is responsible for their violation of Bell's inequality. In the following sections, a probabilistic account of the Bell correlation is presented that uses the facts of classical and quantum optics, and statistics conditional on the phenomena of photon-pair events. On the basis of this calculation, it is concluded that for widely separated optical beams, the formalism of entanglement, though yielding the correct correlation, is sufficient rather than necessary. Thus, the problematic implications regarding nonlocality may ultimately follow from unnecessary excess assumptions contained in the entanglement explanation of Bell correlations.

## 2     Assumptions used in the present model
*Physical facts*

The model to be described rests on experimental phenomena associated with spontaneous parametric down conversion (SPDC) and transient wave-optical interference. The relevant facts of SPDC are described with the help of Fig. 1 which shows photon pairs emanating from a source designed by Kwiat, et al [18]. Due to the phase matching conditions of nonlinear optics for Type II down conversion, two cones of light emanate from a laser pumped crystal, with light of one cone having horizontal polarization and the other having vertical polarization. The photons emitted in this configuration occur in pairs such that the individuals of the pair are found on opposite sides of a ring structure (indicated schematically by the ends of the arrow through the center of the diagram). The process conserves energy and momentum between the pump and output photons, and the two photons of the pair are emitted simultaneously or almost simultaneously [19].

Light from the intersections of the two cones in regions labeled Beam 1 and Beam 2, is selected by apertures in the experimental setup. It is this light that is used to generate Bell correlations. Light from the overlap regions is emitted such that if a Beam 1 photon has horizontal polarization, the corresponding Beam 2 photon has vertical polarization, and vice versa. Nevertheless, from both a wave optics and quantum optics point of view, light from these regions must first be treated as consisting of a superposition of horizontally and vertically polarized waves. The resultant random polarization components along orthogonal analyzer output axes produce correlated wave intensities that correspond to correlated counts from the polarization beam splitters (PBSs) on opposite sides of Bell experiments. Output counts are selected as simultaneous events if two occur in the same observation time interval.

*The probability model*

The calculation of the Bell correlation described below incorporates the experimental facts in a mixed Poisson probability model. The model uses Poisson count probabilities conditionally dependent on exponentially distributed Poisson parameters synonymous with intensity or count averages. The paired H and V complex amplitudes on opposite sides of the apparatus are assumed to have circular complex Gaussian (CCG) statistics [20], resulting in corresponding statistically independent exponentially distributed intensities, and phases that are statistically independent of amplitudes. A condition on the phase of the horizontal (H) relative to the vertical (V) complex amplitudes, to be specified below, results from non-linear optics. So that variables are unit-less, the H and V instantaneous intensity inputs are measured in counts in an observation time (see Appendix A).

The statistics of the inputs to the PBS's are appropriate for "chaotic" light beams [20, 21]. While these statistics determine the number of photons input to the PBS analyzers, the functioning of the PBS's produces output counts assumed to be binomially distributed for fixed input numbers. The input beams are described individually and locally, but have intensity and phase statistical parameters linked at the source. These quantities correspond to Bell's $\lambda$'s [6]. As in the quantum treatment, conditions on statistical parameters result in joint probabilities, but deterministic relations coupling photon behavior to wave statistics (hidden variables) are unknown and (statistically) unnecessary. It should be remembered that local hidden variables for photon behavior have never been specified either quantum mechanically or classically. What have been specified are the statistical count rates of photons based on wave-optical intensities. These statistics are augmented by the empirical fact that photons are not split at beam splitters. This result is applied to the case of indeterminate numbers of photon pairs intrinsic to the model to obtain the correlations of single photon pairs.

## 3     Calculation of the Bell correlation

*Wave optics calculation of field superpositions and corresponding intensities*

The field in the ring overlap region (Fig. 1) of Beam 1 is

$$\vec{U}_1 = u_{1H}\hat{i} + u_{1V}\hat{j}, \qquad (3.1)$$

and of Beam 2 is

$$\vec{U}_2 = u_{2H}\hat{i} + u_{2V}\hat{j}. \qquad (3.2)$$

The u's indicate complex amplitudes of the SPDC output fields in terms of their horizontal (subscript H) and vertical (subscript V) components in regions 1 and 2. Carots indicate unit vectors in the *x* and *y* directions in these and the following equations. Fields $\vec{U}_1$ and $\vec{U}_2$ are incident on PBS's 1 and 2 respectively. Polarizer orientations are indicated by (mutually orthogonal) unit vectors subscripted $n$ and $p$:

$$\hat{n}_{l\text{n}} = \cos\theta_l \hat{i} + \sin\theta_l \hat{j}, \quad \hat{n}_{lp} = -\sin\theta_l \hat{i} + \cos\theta_l \hat{j} \quad l=1,2. \tag{3.3}$$

Unit vector $\hat{n}_{l\text{n}}$ points in the polarizer transmit direction at angle $\theta_l$ counter clockwise from the *x*-axis, and $\hat{n}_{lp}$ points along the reflect direction. Transmitted and reflected complex output amplitudes $U_{l\text{n}}$ and $U_{lp}$ from the two PBS's are then:

$$\begin{aligned} U_{l\text{n}} &\equiv \vec{U}_l \cdot \hat{n}_{l\text{n}} = u_{lH}\cos\theta_l + u_{lV}\sin\theta_l, \\ U_{lp} &\equiv \vec{U}_l \cdot \hat{n}_{lp} = -u_{lH}\sin\theta_l + u_{lV}\cos\theta_l, \quad l=1,2. \end{aligned} \tag{3.4}$$

These output amplitudes are just components of the input vectors along each of the two orthogonal output directions, one pair of outputs for each PBS.

*Calculation of the intensities of the analyzer outputs*

The PBS output intensities in the transmit and reflect directions corresponding to (3.4) are

$$I_{l\text{n}} = U_{l\text{n}}U_{l\text{n}}^*; \quad I_{lp} = U_{lp}U_{lp}^* \quad l=1,2. \tag{3.5}$$

The intensities (3.5) are statistical parameters; they denote instantaneous (number of event) averages, but do not otherwise specify events. From (3.4) and (3.5), the intensities transmitted and reflected, respectively, by the PBS on side 1 are

$$I_{1n} = I_{1H}\cos^2\theta_1 + I_{1V}\sin^2\theta_1 + \sqrt{I_{1H}I_{1V}}\cos(\theta_{1H}-\theta_{1V})\sin 2\theta_1, \tag{3.6}$$

and

$$I_{1p} = I_{1H}\sin^2\theta_1 + I_{1V}\cos^2\theta_1 - \sqrt{I_{1H}I_{1V}}\cos(\theta_{1H}-\theta_{1V})\sin 2\theta_1, \tag{3.7}$$

where $\theta_{1H}$ and $\theta_{1V}$ are the phases of the fields, $u_{1H}$ and $u_{1V}$, respectively, and source intensities are given by $I_{1x} = u_{1x}u_{1x}^*$, $x=H,V$. Intensities $I_{1H}$ and $I_{1V}$ are statistically independent and exponentially distributed with uniformly distributed phases $\theta_{1H}$ and $\theta_{1V}$ that are statistically independent of them.

Similarly, the corresponding intensities on side 2 are

$$I_{2n} = I_{2H} \cos^2 \theta_2 + I_{2V} \sin^2 \theta_2 + \sqrt{I_{2H} I_{2V}} \cos(\theta_{2H} - \theta_{2V}) \sin 2\theta_2, \qquad (3.8)$$

and

$$I_{2p} = I_{2H} \sin^2 \theta_2 + I_{2V} \cos^2 \theta_2 - \sqrt{I_{2H} I_{2V}} \cos(\theta_{2H} - \theta_{2V}) \sin 2\theta_2, \qquad (3.9)$$

where again $\theta_{2H}$ and $\theta_{2V}$ are uniformly distributed and independent of the intensities $I_{2x} = u_{2x} u_{2x}^*$, $x = H, V$.

*Requirements based on the nonlinear optics of SPDC*

The development up to this point employs a standard and quite general complex representation of light of arbitrary polarization to compute output amplitudes of two polarization beam splitters at arbitrary rotational orientations [22]. In equations (3.10) - (3.14) immediately below, accepted facts arising from the quantum and nonlinear optics of SPDC are added to the model. They will then be incorporated into equations (3.6) – (3.9) to characterize Bell experiments.

Photons are emitted from the source at random times in pairs having orthogonal polarizations (Fig. 1). This implies that statistical parameters that specify average counts in an observation time on opposite sides of the ring structure of Fig. 1 (source intensities appearing in (3.6-3.9)) are equal:

$$I_{1H} = I_{2V} \; ; \; I_{1V} = I_{2H}. \qquad (3.10)$$

By adding (3.6) and (3.7), and (3.8) and (3.9), the intensity (count rate) is found to be conserved (unsurprisingly) from input to output for each of the two PBS's if assumed to be lossless

$$\begin{aligned} I_{1n} + I_{1p} &= I_{1H} + I_{1V}, \\ I_{2n} + I_{2p} &= I_{2H} + I_{2V} = I_{1H} + I_{1V}. \end{aligned} \qquad (3.11)$$

*Phase conditions due to nonlinear interactions and waveplate use*

From the nonlinear optics process [23] a phase relation exists between the beams in opposite regions 1 and 2. In addition, a wave plate is used in the optical path [18]. The resulting phase relation is:

$$\theta_{2H} + \theta_{1V} = const + \Delta_{2H},$$
$$\theta_{2V} + \theta_{1H} = const + \Delta_{2V},$$
(3.12)

so

$$\theta_{2H} - \theta_{2V} = \theta_{1H} - \theta_{1V} + \Delta_{2H} - \Delta_{2V}.$$
(3.13)

The external wave plate is set so that

$$\theta_{2H} - \theta_{2V} = \theta_{1H} - \theta_{1V} + \pi$$
(3.14)

When (3.10) and (3.14) are used in (3.8) and (3.9) one obtains

$$I_{2n} = I_{1V} \cos^2 \theta_2 + I_{1H} \sin^2 \theta_2 - \sqrt{I_{1V} I_{1H}} \cos(\theta_{1H} - \theta_{1V}) \sin 2\theta_2,$$
(3.15)

and

$$I_{2p} = I_{1V} \sin^2 \theta_2 + I_{1H} \cos^2 \theta_2 + \sqrt{I_{1V} I_{1H}} \cos(\theta_{1H} - \theta_{1V}) \sin 2\theta_2.$$
(3.16)

For convenience, (3.6) and (3.7) are repeated here as

$$I_{1n} = I_{1H} \cos^2 \theta_1 + I_{1V} \sin^2 \theta_1 + \sqrt{I_{1H} I_{1V}} \cos(\theta_{1H} - \theta_{1V}) \sin 2\theta_1,$$
(3.17)

and

$$I_{1p} = I_{1H} \sin^2 \theta_1 + I_{1V} \cos^2 \theta_1 - \sqrt{I_{1H} I_{1V}} \cos(\theta_{1H} - \theta_{1V}) \sin 2\theta_1.$$
(3.18)

Under these conditions, it is now seen that if $\theta_1$ equals $\theta_2$, then $I_{2n}$ equals $I_{1p}$, and $I_{2p}$ equals $I_{1n}$.

*Computing the joint probabilities for the Bell correlation*

Now that (3.15) - (3.18) have been obtained, their products may be computed and ensemble averaged over exponentially independently distributed input parameters $I_{1H}$, $I_{1V}$, and uniformly and independently distributed phases $\theta_{1H}$ and $\theta_{1V}$. For uniformly distributed phases independent of intensities [20], $\overline{\cos\theta} = 0$, for $\theta = \theta_{1H} - \theta_{1V}$, and (denoting statistical averaging by pointed brackets) one finds for $\langle I_{1n} I_{2p} \rangle$ using (3.16-3.17)

$$\langle I_{1n} I_{2p} \rangle = \overline{I}_{1H}\overline{I}_{1V} \cos^2\theta_1 \sin^2\theta_2 + \overline{I_{1H}^2} \cos^2\theta_1 \cos^2\theta_2 + \overline{I_{1V}^2} \sin^2\theta_1 \sin^2\theta_2 +$$
$$\overline{I}_{1H}\overline{I}_{1V} \sin^2\theta_1 \cos^2\theta_2 + \overline{I}_{1H}\overline{I}_{1V} \overline{\cos^2\theta} \sin 2\theta_1 \sin 2\theta_2. \tag{3.19}$$

For independent exponentially distributed intensity parameters $I_{1H}, I_{1V}$ with mean $\overline{I}_0$, $\overline{I_{jx}^2} = 2\overline{I}_{jx}^2$, $j = 1, 2$, $x =$ H, V, $\overline{I}_{jx} = \overline{I}_0$, and $\overline{\cos^2\theta} = 1/2$. With these substitutions, (3.19) becomes

$$\langle I_{1n} I_{2p} \rangle = \overline{I}_0^2 + 2\overline{I}_0^2 \frac{1}{2} \cos^2(\theta_1 - \theta_2). \tag{3.20}$$

The other correlations of interest may be obtained in a similar manner, and one finds:

$$\langle I_{1n} I_{2p} \rangle = \langle I_{1p} I_{2n} \rangle = \overline{I}_0^2 + 2\overline{I}_0^2 \frac{1}{2} \cos^2(\theta_1 - \theta_2) \tag{3.21a}$$

$$\langle I_{1n} I_{2n} \rangle = \langle I_{1p} I_{2p} \rangle = \overline{I}_0^2 + 2\overline{I}_0^2 \frac{1}{2} \sin^2(\theta_1 - \theta_2), \tag{3.21b}$$

$$\langle I_{1n} I_{1p} \rangle = \langle I_{2n} I_{2p} \rangle = \overline{I}_0^2, \tag{3.21c}$$

The second terms on the right in (3.21a,b) are proportional to the usual quantum probabilities. However, (3.21a-c) also contain a constant offset term. It will be shown below that these offsets are the result of indefinite numbers of photon pairs produced by the chaotic light process, whereas in the conventional quantum treatment only single pairs of photons are assumed a priori.

*Effect of indefinite numbers of photon pairs*

Although the preceding derivation of (3.21a-c) has been described with frequent reference to quantum optical photon-count concepts, the results have been based on a wave-optics formalism that includes the requirements of nonlinear interactions. However, in order to account for the offset constants, it is necessary to interpret the ensemble average results of (3.21c) in terms of the statistics of photon counts. The result will then be applied to (3.21a,b) as well.

A key empirical quantum mechanical fact must be added to the nonlinear optics constraints on the statistics: individual photons are not split by a beam splitter/polarization analyzer, but after incidence are emitted in a mutually exclusive manner from one output port or the other. For example, in the case of analyzer 1, if the sum of the H and V input photons is $n_{HV}$, the numbers emitted at the output ports are $n_{1n}$ and $n_{1p} = n_{HV} - n_{1n}$. Using $E\{\cdot\}$ to denote averaging, the correlation of these output numbers is conditional on $n_{HV}$, that is

$$E\{n_{1n}n_{1p} \mid n_{HV}\} = E\{n_{1n}(n_{HV} - n_{1n})\} = E\{n_{1n}n_{HV} - n_{1n}^2\}. \quad (3.22)$$

(The number of photons must be conserved for a lossless analyzer.) Using $p_{1n}$ for the binomial probability that an incident photon is emitted from port $1n$, (3.22) becomes

$$E\{n_{1n}n_{1p} \mid n_{HV}\} = E\{n_{1n}n_{HV} - n_{1n}^2\} = p_{1n}n_{HV}^2 - [n_{HV}(n_{HV} - 1)p_{1n}^2 + n_{HV}p_{1n}]$$
$$= p_{1n}p_{1p}[n_{HV}^2 - n_{HV}], \quad (3.23)$$

where $p_{1n} + p_{1p} = 1$. (This has exactly the same form as the correlation of the number-of-heads $\times$ number-of-tails for $n_{HV}$ flips of a coin given a probability of heads equal to $p_{1n}$.)

The total number of input photons $n_{HV}$ is the sum of contributions from a pair of independent Poisson processes conditional on the two exponentially distributed wave intensities $I_{1H}$ and $I_{1V}$, respectively. Relation (3.23) must now be Poisson averaged using Poisson parameter $\bar{n}_{HV}$, $\bar{n}_{HV} = I_{1H} + I_{1V}$, with $p_{1n}$ and $p_{1p}$ defined from (3.17) and (3.18), respectively, divided by the total input intensity $\bar{n}_{HV}$. (Note: the Poisson parameter due to the sum of two independent Poisson processes is the sum of the Poisson parameters.) Thus, averaging (3.23) using Poisson density $P_{Poisson}(n_{HV} \mid \bar{n}_{HV})$ yields

$$E_{Poisson}\{E\{n_{1n}n_{1p} \mid n_{HV}\}\} = p_{1n}p_{1p} \sum_{n_{HV}=0}^{\infty} (n_{HV}^2 - n_{HV}) P_{Poisson}(n_{HV} \mid \bar{n}_{HV})$$
$$= p_{1n}p_{1p} \sum_{n_{HV}=2}^{\infty} (n_{HV}^2 - n_{HV}) P_{Poisson}(n_{HV} \mid \bar{n}_{HV}) = p_{1n}p_{1p}[\bar{n}_{HV}(\bar{n}_{HV} + 1) - \bar{n}_{HV}] = p_{1n}p_{1p}\bar{n}_{HV}^2. \quad (3.24)$$

It is important to note that the sum in (3.24) depends only on terms for which $n_{HV} \geq 2$. If no photons are incident on the polarization beam splitter, the output from each port is zero, and there is no contribution to the correlation. Similarly, if one photon is incident, the correlation at the outputs is zero (since one photon cannot be split) and that contribution to the sum is zero. Thus, it is the intrinsically indeterminate number ($n_{HV} \geq 2$) of input photons that characterizes chaotic light that causes a nonzero result in (3.24), even if $\bar{n}_{HV}$ is small.

Given $I_{1n}$ and $I_{1p}$ from (3.17-3.18), and $p_{1n}$ and $p_{1p}$ equal to $p_{1n} = I_{1n}/\bar{n}_{HV}$ and $p_{1p} = I_{1p}/\bar{n}_{HV}$, (3.24) equals

$$E_{Poisson}\{E\{n_{1n}n_{1p} | n_{HV}\}\} = p_{1n}p_{1p}\bar{n}_{HV}^2 = \frac{I_{1n}}{(I_{1H}+I_{1V})}\frac{I_{1p}}{(I_{1H}+I_{1V})}(I_{1H}+I_{1V})^2 \quad (3.25)$$
$$= I_{1n}I_{1p}.$$

Finally, (3.25) must be averaged with respect to $I_{1H}$, $I_{1V}$, and $\theta$ to obtain $\bar{I}_0^2$ in (3.21c). Using (3.25), the complete calculation is then

$$\iint_{I_{1H}I_{1V}}\int_\theta dI_{1H}\, dI_{1V}\, d\theta P(I_{1H})P(I_{1V})P(\theta)E_{Poisson}\{E\{n_{1n}n_{1p}|n_{HV}\}\}$$
$$= \iint_{I_{1H}I_{1V}}\int_\theta dI_{1H}\, dI_{1V}\, d\theta P(I_{1H})P(I_{1V})P(\theta)I_{1n}I_{1p} = \bar{I}_0^2. \quad (3.26)$$

An identical analysis yields the same result for $\langle I_{2n}I_{2p}\rangle$.

Thus, the possibility of randomly occurring input numbers greater than 1, inherent in the probability densities describing the physical processes, leads to constant offsets even if $\bar{I}_0$ is small. By contrast, in the quantum mechanical treatment, it is assumed a priori, that only one photon at a time is incident on each analyzer. To relate the case $n_{HV} = 0, 1, 2, \cdots$ to the quantum treatment, it is necessary to subtract $\bar{I}_0^2$ from the average of (3.21c).

One may now use the fact that for $\theta_1 = \theta_2 = \theta$, $I_{1n}(\theta) = I_{2p}(\theta)$. Then $\bar{I}_0^2$ should be subtracted from both sides of

$$\langle I_{1n}(\theta)I_{1p}(\theta)\rangle = \langle I_{1p}(\theta)I_{2p}(\theta)\rangle \quad (3.27a)$$

to convert from an indefinite Poisson input number to an a priori single photon input. The condition for $\theta_1 = \theta_2 = \theta$ determines that the same constant must be subtracted for $\theta_1 \neq \theta_2$. Similarly, since $I_{2n}(\theta) = I_{1p}(\theta)$, the correlations described by the relation

$$\langle I_{1n}(\theta)I_{1p}(\theta)\rangle = \langle I_{1n}(\theta)I_{2n}(\theta)\rangle, \tag{3.27b}$$

are made consistent by subtracting $\bar{I}_0^2$ from both sides. Thus, removing the constant $\bar{I}_0^2$ from (3.21b) yields the correlation for single photon pairs based on the statistical model used.

The same condition holds for the two remaining correlations (3.21a). From (3.15) - (3.18)

$$\begin{aligned} I_{2p}(\theta + \pi/2) &= I_{2n}(\theta) = I_{1p}(\theta) \\ I_{2n}(\theta + \pi/2) &= I_{2p}(\theta) = I_{1n}(\theta). \end{aligned} \tag{3.27c}$$

Applying this to (3.21a) yields

$$\begin{aligned} \langle I_{1n}(\theta)I_{2p}(\theta + \pi/2)\rangle &= \langle I_{1n}(\theta)I_{1p}(\theta)\rangle = 0 \\ \langle I_{1p}(\theta)I_{2n}(\theta + \pi/2)\rangle &= \langle I_{1n}(\theta)I_{1p}(\theta)\rangle = 0. \end{aligned} \tag{3.27d}$$

Thus, subtraction of $\bar{I}_0^2$ from (3.21c) requires it to be subtracted from the correlations of (3.21a) as well.

The subtraction of $\bar{I}_0^2$ resulting from (3.27) also leads to recovery of the single pair boundary conditions at $\theta = 0$ stated at the beginning of the discussion.

$$\langle I_{1n}(\theta)I_{2n}(\theta)\rangle = \langle I_{1n}(\theta)I_{1p}(\theta)\rangle = 0$$

implies that

$$\langle I_{1n}(0)I_{2n}(0)\rangle = \langle I_{1n}(0)I_{1p}(0)\rangle = 0,$$

and

$$\langle I_{1p}(\theta)I_{2p}(\theta)\rangle = \langle I_{1n}(\theta)I_{1p}(\theta)\rangle = 0$$

implies

$$\langle I_{1p}(0)I_{2p}(0)\rangle = \langle I_{1n}(0)I_{1p}(0)\rangle = 0.$$

The final conclusion is that the wave-optical results (3.21a-c) are consistent with the quantum mechanical relations when intensities are interpreted in terms of photon count rates, and the contribution of an intrinsically indeterminate number of input photons is subtracted out.

*Computation of the Bell correlation*

Using (3.21a,b), the Bell correlation may be computed from the average with respect to $I_{1H}$, $I_{1V}$, and $\theta$ as indicated by the subscripted pointed bracket

$$C = \langle I_{1n}I_{2n} + I_{1p}I_{2p} - I_{1p}I_{2n} - I_{1n}I_{2p} \rangle_{HV\theta} = -2\overline{I}_0^2 \cos 2(\theta_2 - \theta_1). \qquad (3.23)$$

Equations (3.21a,b) must be corrected for single photon pair input to compute the normalization sum

$$\langle I_{1n}I_{2n} + I_{1p}I_{2p} + I_{1p}I_{2n} + I_{1n}I_{2p} - 4\overline{I}_0^2 \rangle_{HV\theta} = 2\overline{I}_0^2. \qquad (3.24)$$

The normalized expression for the Bell correlation C is then (3.23) divided by (3.24) or

$$C_{normalized} = -\cos 2(\theta_2 - \theta_1). \qquad (3.25)$$

**4.    Conclusion**

A local probability calculation has been presented that results in the Bell correlation. In Bell's book [24], he indicated that an expression analogous to (3.21) (corrected for single photon input) could not be used to obtain the Bell correlation since the Bell inequality would then be violated. However, in the introduction, it was pointed out that the reasoning used in the Bell theorem does not exclude local models built on intrinsically non-commutative processes such as used here. If a local probability model yields the same correlation as the nonlocal probability model based on entanglement, then it must be concluded that the Bell correlation is not a unique result of the entanglement model.

While the formal probability model given yields the Bell correlation, a more detailed interpretation of the relation of the ensemble averages used to the ultimate occurrence of photon events would be desirable. The situation here is rather analogous to that of quantum mechanics: the photon events follow wave intensities and their correlations, but the formalism is reticent as to how this occurs. In the present model, ensemble averages must be performed before the quantum mechanical probability is obtained. Thus the underlying statistical processes on which the calculation depends appear as "hidden" processes. Since photon coincidence count observations depend on an ensemble of values of underlying random variables, there is a possibility of a wave packet description of the photon as occurs in the quantum description [25]. The correlation of photon occurrences would again equal appropriate correlations of wave packet variables computed through ensemble averages.

ACKNOWLEDGMENTS

The author is grateful to Michael Steiner, Ronald Rendell, Armen Gulian, and Xiaolei Zhang while at the Center for Quantum Studies at George Mason University (GMU) for many stimulating discussions and critical comments on the manuscript, and Joe G. Foreman for critiques of the presentation. He would also like to thank Jeff Tollakson at the Center for Quantum Studies at GMU for his kind hospitality during preparation of much of the manuscript.

**Appendix A**

A number of studies indicate that the outputs of a spontaneous parametric down converter (SPDC) are twin beams for which each beam individually exhibits thermal statistics matched to the other beam. Thus, one may associate with each beam the same fluctuating intensity $I$ (measured as counts in a fixed observation time) having probability density

$$P(I) = \frac{1}{I_0} e^{-I/I_0}. \tag{A1}$$

The count/intensity $I$ serves as the Poisson parameter of a Poisson count probability density providing pairs of matched count numbers, the same number for each beam. The Poisson probability is conditional on the intensity (average count number) and is given by:
$$P(n|I) = \frac{e^{-I} I^n}{n!}. \tag{A2}$$

One may then use (A1) and (A2) to define a joint density $P(n,I)$ as

$$P(n,I) = P(n|I)P(I). \tag{A3}$$

In Bell experiments, events are selected in which a single count occurs simultaneously on the two sides of the experiment.

*P(n)* may be computed as the integral over *I* in (A3) to yield:

$$P(n) = \frac{\bar{I}_0^n}{(\bar{I}_0 + 1)^{n+1}}.$$  (A4)

The resulting average of *n* in turn is then

$$\sum_{0}^{\infty} n \frac{\bar{I}_0^n}{(\bar{I}_0 + 1)^{n+1}} = \bar{I}_0$$

FIGURE

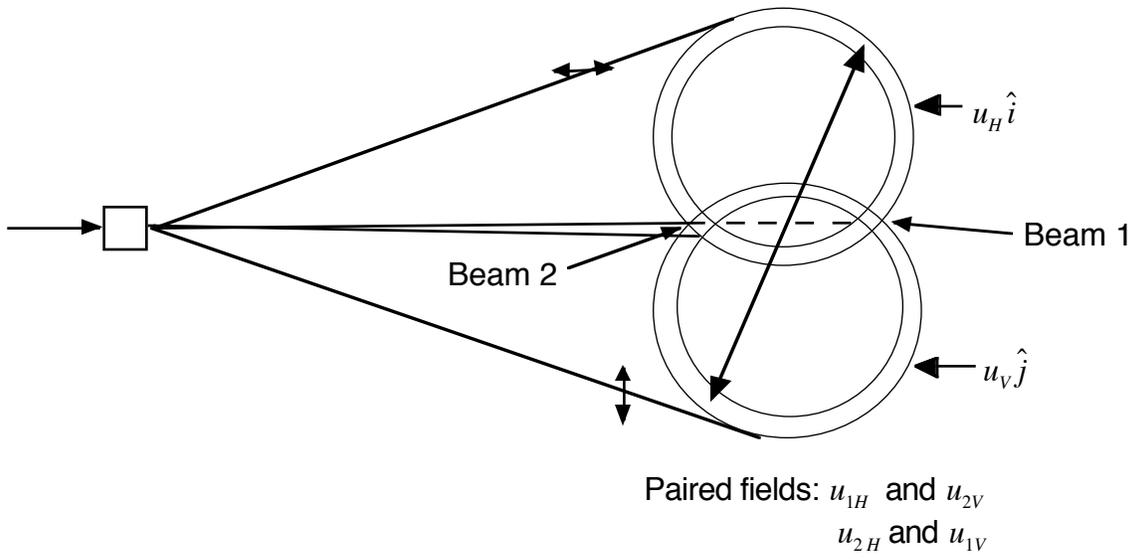

Fig. 1, Output of an SPDC in type 2 configuration.